\documentclass[10pt,journal]{IEEEtran}
\usepackage{amsmath,amssymb,epsfig,cite,footnote,authblk}
\usepackage{color,hyperref}
\hypersetup{
    colorlinks,%
    citecolor=blue,%
    filecolor=blue,%
    linkcolor=blue,%
    urlcolor=blue
}
\definecolor{bluecolor}{rgb}{0,0.,1.}

\definecolor{redcolor}{rgb}{.7,0.,0.}

\newcommand{\pr}[1]{\left( #1\right)}
\newcommand{\prr}[1]{\left[ #1 \right]}
\newcommand{\es}[1]{\begin{equation}\begin{split}#1\end{split}\end{equation}}
\newcommand{\est}[1]{\begin{equation*}\begin{split}#1\end{split}\end{equation*}}

\newcommand{\R}{\mathbb{R}}

\newcommand{\V}{\mathcal{V}}

\newcommand{\rr}{\mathbf{r}}
\newcommand{\dd}{\textrm{d}}

\begin{document}

\title{Multihop connectivity of ad hoc networks with randomly oriented directional antennas}
%\author[1]{2 authors}
\author[1]{Orestis Georgiou}
\author[2]{Camly Nguyen}
\affil[1]{Toshiba Telecommunications Research Laboratory, 32 Queens Square, Bristol, BS1 4ND, UK}
\affil[2]{Network System Laboratory, Corporate Research \& Development Center, Toshiba Corporation,
1 Komukai-Toshiba-cho, Saiwai-ku, Kawasaki 212-8582, Japan}
\maketitle

%\pagestyle{plain}
%%%%%%%%%%%%%%%%%%%%%%
%\thispagestyle{fancy}

\begin{abstract}
Directional antennas and beamforming can significantly improve point-to-point wireless links when perfectly aligned.
In this letter we investigate the extreme opposite where antenna orientations and positions are chosen at random in the presence of Rayleigh fading.
We show that while the 1-hop network connectivity is deteriorated, the multihop routes improve, especially in the dense regime.
We derive closed form expressions for the expectation of the $1$-hop and $2$-hop degree which are verified through computer simulations.
We conclude that node density does not greatly affect the number of hops required between stations whilst simple random beamforming schemes do, thus returning substantial network performance benefits due to the existence of shorter multi-hop paths.
\end{abstract}

%\begin{keywords}
%Wireless multihop networks, anisotropic radiation, directivity, connectivity.
%\end{keywords}

%%%%%%%%%%%%%%%%%%%%%
\section{Introduction \label{sec:intro}}

Wireless ad hoc networks are typically equipped with multihop relaying and signal processing capabilities and find application in sensor and mobile systems such as smart grid, industrial and environmental monitoring, search-and-rescue operations etc.
Commonality in many of these applications arises in that the number and distribution of nodes in the networks is often random motivating the study of \textit{random geometric graphs} \cite{penrose2003random}.
In the simplest case, these graphs consist of a large number of points scattered in a region of space and paired up whenever their mutual distance is less than some scalar value $r_0$.
A plethora of generalizations of this basic model have been put forward in an attempt to understand the connectivity properties of ad hoc networks and suggest improved network design and deployment methodologies \cite{younis2008strategies}.

In random graph models, one is typically interested in statistical properties such as the \textit{hop distribution} portraying the multi-hop connectivity as popularized in the theatrical play ``Six Degrees of Separation''.
In wireless ad hoc networks, the hop distribution was originally investigated by Chandler \cite{chandler1989calculation} who concluded that ``The node density does not greatly affect the number of hops required between two nodes, but has a much greater effect on whether a connection can be made at all''.
Recently, a recursive formula for calculating the probability that a pair of nodes ($i,j$) separated by a distance $r_{ij}$ is connected in at least $k$-hops was given in \cite{mao2010probability}.
The effects of randomly oriented directional (or anisotropic) antennas or random beamforming schemes (where nodes  beamform in a random direction) were numerically studied in\cite{vilzmann2005hop,zhou2009connectivity} suggesting that such simple schemes can lead to network performance gains in routing, end-to-end delay, reachability, interference tolerance, and capacity.
Recent analytical work has shown that the reported benefits are actually highly dependent on the prevailing channel conditions described by path loss exponent $\eta$ experienced in the propagation medium \cite{coonanisotrop,georgiou2013connectivity}.
%These are however restricted to just $1$-hop statistics.

\begin{figure}[t]
\centering
\includegraphics[scale=0.255]{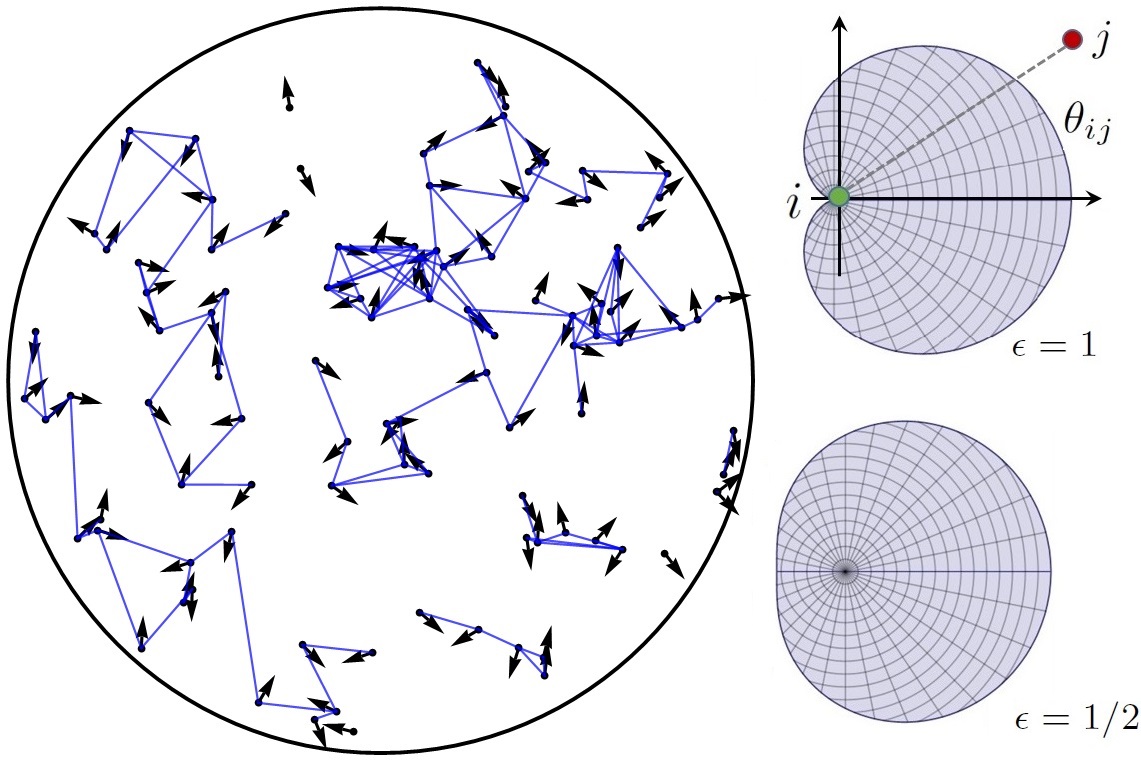}
\caption{\textit{Left:} Example network realization with $N=100$ random directional nodes in a disk domain, using $\rho\!=\!\beta\!=\!\epsilon\!=\!1$ and $\eta\!=\!4$. The blue arrows show the directional links available that can be utilised for routing and scheduling transmissions. \textit{Right:} Cardioid radiation patterns for $\epsilon=1/2,1$.}
\label{fig:illust}
\end{figure}

In this letter we analytically quantify these benefits for the first time and obtain closed form expressions for the multihop reachability of random networks with randomly oriented antennas in the presence of Rayleigh fading.
We therefore show when and how schemes like random beamforming provide progressively better multihop coverage and reachability (measured by the number of accessible nodes in at least $k$ hops).
This follows from the more likely existence of shorter multi-hop paths between stations in the dense network regime with randomly oriented directional antennas.
Implicit for this performance improvement is perfect interference management and a good MAC.
We derive communication theoretic lower bounds on the multihop connectivity of networks with randomly oriented directional antennas, useful when designing wireless sensor networks (WSNs), for instance, for choosing the right density of directional nodes to be deployed in order to meet certain multihop requirements or performance goals.
Significantly, we also confirm Chandler's conclusion \cite{chandler1989calculation} by showing that the typical hop distance $\bar{h}\sim \mathbf{c} \rho^{-1/2}$ where the coefficient $\mathbf{c}$ strongly depends on the path loss exponent $\eta$ and antenna directivity.
The exact form of $\mathbf{c}$ remains an interesting open question.

%%%%%%%%%%%%%%%%%%%%%
\section{Network Definitions and System Model \label{sec:model}}

Consider a wireless network with identical nodes located randomly on the plane $\R^2$ according to a spatial Poisson point process (PPP) of intensity $\rho$. We denote the location of node $i$ by the vector $\rr_i=(r_i,\theta_i)$ in polar coordinates.
Furthermore, consider the case where each node is equipped with a directional antenna pointing in a direction $\vartheta_i$ chosen randomly in $[0,2\pi]$.
Such a configuration is commonly found in WSN applications where sensors or smart meters form a random mesh topology (see for example Fig. \ref{fig:illust}).

Assuming negligible inter-node interference (e.g. perfect CDMA/TDMA), we define the connection probability between transmitting node $i$ and receiving node $j$ through the relation
$H_{ij}=\mathbb{P}(\textrm{SNR} \cdot |h|^2 \geq \wp)$,
where SNR denotes the long-term average received signal-to-noise ratio and $h$ is the channel transfer coefficient for single input single output (SISO) antenna systems. 
Assuming identical lossless antennas we have from the Friis transmission formula that
$
\textrm{SNR} \propto G_{ij} G_{ji}  r_{ij}^{-\eta}
$,
where $r_{ij}=|\rr_i - \rr_j|$ is the Euclidean distance between the two nodes, $\eta$ is the path loss exponent (typically $\eta\geq2$), and
$G_{ij}$ is the gain of the antenna at node $i$ observed in the direction of node $j$.
Notice that $H_{ij}=H_{ji}$ since the channel is assumed reciprocal.
Isotropic antennas have a constant gain $G=1$, while anisotropic ones are functions of the polar angle $\theta$, appropriately normalized by the condition $\int_{0}^{2\pi } G(\theta) \dd \theta =2\pi$.
Assume that the antenna gain is given by a cardiod function 
\es{
G_{ij}( \theta_{ij})=1 +\epsilon \cos \theta_{ij},
\label{cc}}
where $\epsilon\in[0,1]$ measures the extent of deformation from the isotropic case ($\epsilon=0$), and $\theta_{ij}$ is the direction of receiving node $j$ relative to the antenna orientation of node $i$ (see Fig. \ref{fig:illust}).
The cardiod function \eqref{cc} is representative of wide-angle unidirectional radiation patterns often found in microstrip (patch) antennas.
We expect however that the results presented herein qualitatively apply to other directional gain functions as well as multi-directional e.g. $G= 1+\epsilon \cos n \theta$ for $n>1$ as was the case in previous studies \cite{georgiou2013connectivity}.

The framework that follows is statistical and assumes a Rayleigh fading channel such that $|h|^2 \sim \exp(1)$ and
\es{
H_{ij}(r_{ij})= \mathbb{P}\pr{ |h|^2 \geq -\wp/\textrm{SNR} } = e^{ -\beta r_{ij}^{\eta}/ (G_{ij} G_{ji} )}
,
\label{HH}}
where $\beta$ depends on the transmission wavelength, signal power, the threshold $\wp$ etc., and defines an effective communication range $r_0= \prr{G_{ij} G_{ji}/\beta}^{1/\eta}$.
In simulations we set $\beta=1$.
Connection probability functions like \eqref{HH} have been used extensively in the literature however one should note that there are many other possible functions (e.g. including shadowing or a different path loss function) and it is not clear which one models real networks more accurately.
Equations \eqref{cc} and \eqref{HH} are therefore examples used for the sake of juxtaposition with some of our results and approach being more general.

%%%%%%%%%%%%%%%%%%%%%%%%%%%%%%%%%%%%

\section{K-hop Connectivity \label{sec:khop}}

\subsection{Mean 1-hop degree}
It is often desirable in WSNs that broadcasts are heard by as many neighbouring nodes as possible.
A simple way of measuring this is the mean degree.
%; often called the connectivity of the network.
Denoting the $1$-hop degree of transmitting node $i$ by $d_i^{(1)}$, the $1$-hop mean degree $\mu_1$ is given by the arithmetic mean of $d_i^{(1)}$.
Equivalently, since nodes are Poisson (i.e. uniformly and independently) distributed, we take the spatial and orientation average of equation \eqref{HH} multiplied by the density and normalised by $2\pi$ and calculate
\es{ 
\mu_1  
%=\frac{\rho}{2\pi } \! \int \!\int\!  H_{ij} \, \dd \rr_j \dd \vartheta_j
&= \frac{\rho}{2\pi} \! \int_{0}^{2\pi}\!\! \int_{0}^{2\pi}\!\!\int_0^\infty  \!\! H_{ij}(r_j) r_j \dd r_j\dd\theta_j\dd\vartheta_j \\
%&\approx \frac{\rho}{2\pi}  \! \int_{0}^{2\pi} \!\! \int_{0}^{2\pi} \!\! \int_0^\infty \!\! r e^{- \frac{\beta r^\eta}{(1+\epsilon \cos \theta)(1+\epsilon \cos (\pi-\vartheta+\theta))} }  \dd r \dd \theta\dd \vartheta \\
%= \frac{\rho \Gamma(2/\eta)}{2\pi\eta \beta^{2/\eta}} \int_{0}^{2\pi} \!\! \int_{0}^{2\pi} \!\! (G_{ij} G_{ji})^{2/\eta}  \dd \theta_j \dd \vartheta_j \\
&= \frac{\rho \Gamma(2/\eta)}{2\pi\eta \beta^{2/\eta}} \pr{ \int_{0}^{2\pi} \!\! G^{2/\eta}  \dd \vartheta }^2
\label{mu01}.
}
In the first line of \eqref{mu01} we have used the homogeneity of the spatial PPP on $\R^2$ and set node $i$ at the coordinate origin, oriented along the $x$-axis.
Therefore, in \eqref{mu01} the first two integrals ($\dd r_j\dd\theta_j$) average over all possible positions of $\rr_j \in \R^2$, and the third integral ($\dd\vartheta_j$) averages over all possible orientations $\vartheta_j\in [0,2\pi]$ of antenna $j$. 
In the last line of \eqref{mu01} we have used the periodicity property of the cardiod function
\es{
\int_{0}^{2\pi} \!\! G^{2/\eta}  \dd \theta &=  \pi \Big[ (1-\epsilon)^{\frac{2}{\eta}} {}_2 F_1\pr{\frac{1}{2},-\frac{2}{\eta},1,\frac{2\epsilon}{\epsilon-1}}  \\
&+ (1+\epsilon)^{\frac{2}{\eta}} {}_2 F_1\pr{\frac{1}{2},-\frac{2}{\eta},1,\frac{2\epsilon}{\epsilon+1}} \Big]
%&= 2\pi - \frac{\pi(\eta-2)\epsilon^2}{\eta^2} - \mathcal{O}(\epsilon^4) 
\label{funct}
,}
where ${}_2 F_1$ is the Gauss hypergeometric function.
%Setting $\beta=1$, equation \eqref{mu01} has a minimum at $\eta=4.7334$ when $\epsilon=1$ and at $\eta=4.3325$ when $\epsilon=0$.
For any $\eta>2$, $\mu_1$ is monotonically decreasing with $\epsilon$. 
This implies that \textit{the more directional the antenna gain is, the less nodes can be reached in a single hop}.
Note that since we have assumed identical antennas for both transmitter and receiver chains (i.e. $H_{ij}=H_{ji}$), the in- and out-degree of network nodes are equal.
%Note that this qualitative picture can be shown to hold for isotropic-anisotropic, and anisotropic-isotropic transmitter-receiver pairs.

%%%%%%%%%%%

\subsection{Mean 2-hop degree}
The $1$-hop mean degree $\mu_1$ is a good measure of the local connectivity of a network. 
Often however, and particularly in WSN MAC design \cite{bachir2010mac}, it is desirable to maximize the number of $2$-hop neighbours that a broadcast can reach.
Let the $2$-hop degree of transmitting node $i$ be denoted by $d_i^{(2)}$ counting the number of $2$-hop neighbours of node $i$ which are not $1$-hop neighbours. 
To make progress, define the probability of nodes $i$ and $j$ having at least one common $1$-hop neighbour as $H_{ij}^{(2)}$ which can be expressed as the complement of not having any
\es{
H_{ij}^{(2)}= 1- \prod_{k\not= i,j} (1- H_{ik} H_{kj})
\label{H2}
,}
where nodes $k\not=i,j$ act as a relays. 
Using \eqref{H2} and assuming that $|\rr_i|=\vartheta_i=0$ we may express the $2$-hop mean degree as
\es{
\mu_2 \! &= \frac{\rho}{2\pi} \! \int_0^{2\pi} \!\int_0^{2\pi}\! \int_0^\infty \!(1-H_{ij}) \langle H_{ij}^{(2)} \rangle  r_j \, \dd r_j \dd \theta_j \dd \vartheta_j  
,\label{mu2}}
%where the average of $H_{ij}^{(2)}$ is given by
\es{
\langle H_{ij}^{(2)} \rangle \!\!
%&= 1 - \prod_{k\not= i,j} \pr{ 1-\frac{1}{2\pi V } \!\int_{\M} \!\! H_{ik} H_{kj} \, \dd \rr_k \dd \vartheta_k } \\
%&= 1- \Big( 1  -  \frac{1}{2\pi V} \int_\mathcal{M}\!\! H_{ik} H_{kj} \, \dd \rr_k \dd \vartheta_k\Big)^{N-2} \\
&= \! 1 - \exp\pr{-\frac{\rho}{2\pi} \! \int_0^{2\pi}\!\!\int_{\R^2} \!\! H_{ik} H_{kj} r_k \, \dd r_k \dd \theta_k \dd \vartheta_k}
\label{HH2}}
It is therefore immediately clear that $\mu_2$ is not a linear function of the density $\rho$ (c.f. \eqref{mu01}).
For example, for a simplified unit disk model (i.e. ignoring fading) it can be shown (see Appendix \ref{sec:ap1}.) that $\mu_2 \sim 3\rho\pi r_{0}^2 - 2\pi(2 r_{0})^{2/3} \Gamma(2/3) (\rho/3)^{1/3}$. 
Does the mean $2$-hop degree $\mu_2$ increase or decrease with antenna directionality $\epsilon$? 
This cannot be answered by simply looking at equation \eqref{mu2} and therefore we postpone further discussion for the next section where we numerically evaluate the integrals and compare with computer simulations.

%%%%%%%%%%%%%%%%%
\subsection{Mean k-hop degree}
Generalizing \eqref{mu2} to the case of the $k$-hop mean degree is not straight forward.
Let the $k$-hop degree of transmitting node $i$ be denoted by $d_i^{(k)}$ counting the number of $k$-hop neighbours of node $i$ which are not $(k-1)$-hop. 
The $k$-hop mean degree $\mu_k$ is therefore given by the arithmetic mean of $d_i^{(k)}$.
Define the probability that nodes $i$ and $j$ have at least one common $m$-hop neighbour, which with some care can be expressed as
\es{
H_{ij}^{(m)}= 1- \prod_{k\not= i,j} \prr{ 1- H_{ik}^{(m-1)} H_{kj}^{(1)} \prod_{n=1}^{m-2}\pr{1- H_{ik}^{(n)}} }
\label{Hm}}
where we have used $H_{ij}^{(1)}=H_{ij}$ for ease of notation.
Equation \eqref{Hm} is a nested equation calculating the complement probability of two nodes having no common $m$-hop neighbour, which is itself given by a similar expression.
The $k$-hop mean degree is therefore given by $\frac{\rho}{2\pi}$ times the expected probabilities of two nodes being $k$-hop neighbours but not $m$-hop, for $m \leq k-1$
\es{
\mu_k \! = \! \frac{\rho}{(2\pi)^{N}} \!\int_{\mathcal{M}^N}\!\!\!\! H_{ij}^{(k)}\!\prod_{m=1}^{k-1} \! \pr{\! 1\! -\! H_{ij}^{(m)}\!} \dd \rr_1 \dd \vartheta_1 \!\ldots\! \dd \rr_N \dd \vartheta_N
\label{muk}
}
in the limit of the number of nodes $N\to\infty$ and where we have defined $\mathcal{M}=\R^2 \times [0,2\pi]$.
Equation \eqref{muk} is not particularly insightful and cannot be easily computed, even numerically, or even after some approximations \cite{mao2010probability}.
In analogy to the result in Appendix \ref{sec:ap1}. one may expect that $\mu_k = \rho (2k -1) \pi r_0^2 - \mathcal{O}(\rho^{1/3})$. 
We will confirm this in Sec. \ref{sec:NumAna} for the case of $k=3$, however are unsure of the correction term for higher values of $k$.

%%%%%%%%%%%%%%%%%%%%%

\section{The Typical Hop Distance \label{sec:khopdistrib}}

Given a large but finite network of $N$ nodes, pick two nodes at random.
What is the expected hop distance $\bar{h}$ between these two nodes?
Numerical studies have shown that random beamforming schemes can reduce this number in mesh networks, thus resulting in faster message dissemination while also reducing the total signal processing done by relays \cite{vilzmann2005hop}.
To provide a formal definition of the typical hop distance we begin from the identity stating that the sum of all $k$-hop neighbours of any node $i$ equals to $N-1$. Symbolically this is:
$
 d_i^{(\infty)} + \sum_{k=1}^{N-1} d_i^{(k)} \equiv N-1
%\label{triv}
$,
where $d_i^{(\infty)}$ is the number of nodes which cannot be reached by $i$ in any number of hops. 
Note that $1+\sum_{k=1}^{N-1} d_i^{(k)}$ is the size of the connected component that node $i$ belongs to.
In fact, averaging this over all nodes gives the typical cluster size in the network.
Clearly in a fully connected network $d_i^{(\infty)}=0, \,\forall\, i$.
Averaging both sides of the identity over $i$ and changing the summation order 
\es{
\frac{1}{N}\sum_{i=1}^{N} \pr{ d_i^{(\infty)} + \sum_{k=1}^{N-1} d_i^{(k)} }  = \mu_\infty +  \sum_{k=1}^{N-1} \mu_k \equiv N-1
\label{triv3}
,}
indicating that $\mu_k /(N-1)$ is the probability mass function of the hop distribution for large random networks.
Significantly, the typical hop distance $\bar{h}$ between a random pair of nodes is given by the mean of this hop distribution
$
 \bar{h} = \frac{1}{N-1}\sum_{k=1}^{\infty} k \mu_k 
$.
We have therefore established a link between the typical hop distance $\bar{h}$ and the $k$-hop mean degree \eqref{muk}.
Notice that if the network is not fully connected (i.e. $\mu_\infty \not=0$) then $\bar{h}$ diverges.

%%%%%%%%%%%%%%%%%%%%%%%%%
\section{Numerical Analysis \label{sec:NumAna}}

\begin{figure}[t]
\centering
\includegraphics[scale=0.165]{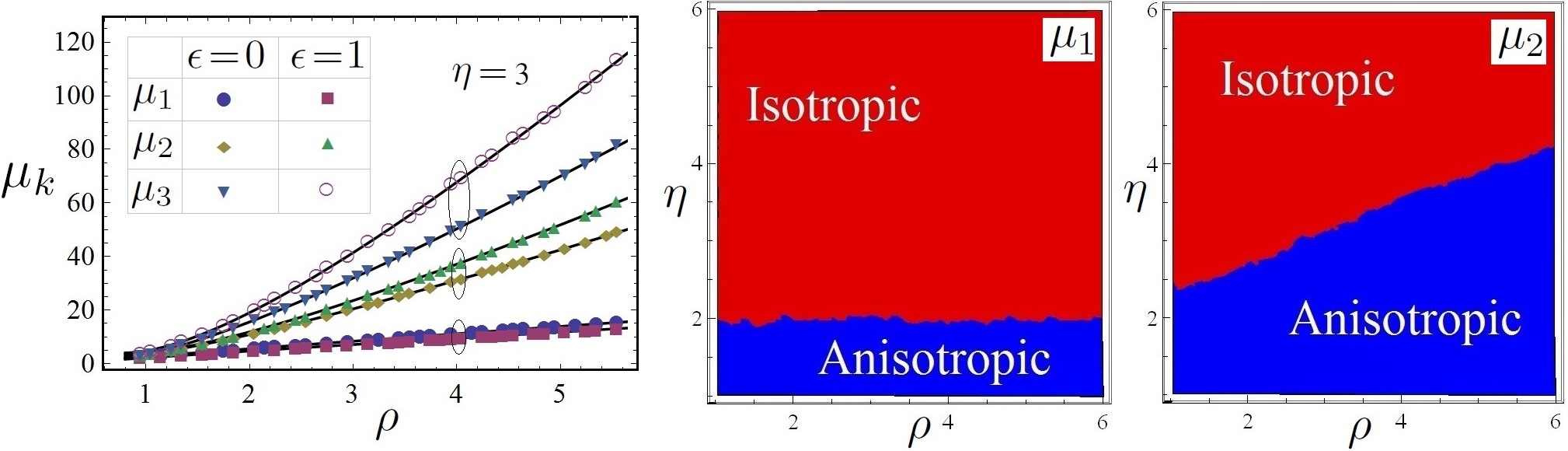}
\caption{\textit{Left:} Simulation results for the $k$-hop mean degree $\mu_k$ as a function of the density $\rho$ for the isotropic ($\epsilon=0$) and anisotropic ($\epsilon=1$) case using $\eta=3$.
For $k=1$, the two black straight lines are given by \eqref{mu01}. For 
$k=2$ the black curves are obtained by numerically integrating \eqref{mu2}, while for $k=3,$ the black curves are fitted to $a-b\rho^{1/3}+c\rho$ for $a,b,c>0$.
The markers are obtained from computer simulations in a large network similar to that in Fig. \ref{fig:illust}.
\textit{Middle and Right:} Parameter space ($\rho,\eta$) showing isotropic (red) or anisotropic (blue) radiation patterns deliver higher $\mu_{1}$ and $\mu_2$ respectively.
}
\label{fig:iso}
\end{figure}

In the simulations that follow, we independently and uniformly generate the spatial and orientation coordinates of $N=\lfloor\rho V\rfloor$ nodes in a circular domain $\V$ of area $V=\pi R^2$ and radius $R=10$ using $\beta=1$. 
Whenever a randomly generated number $\zeta\in[0,1] \leq H_{ij}$, nodes $i$ and $j$ are paired up (see Fig. \ref{fig:illust}).
This guarantees that the network links are statistically independent.
The resulting undirected graph connections are stored in a symmetric adjacency matrix from which we extract useful network observables such as the $k$-hop degree of each node.
In order to keep our results free of boundary effects, we use values drawn from nodes situated away from the border of $\V$ as done in \cite{bettstetter2005does}.
Finally, we improve our statistics by repeating the above process in a Monte Carlo fashion.

In the left panel of Fig. \ref{fig:iso} we plot the simulation results for $\mu_k$ for $k=1,2,3,$ as functions of the node density $\rho$ for $\eta=3$.
$\mu_1$ increases linearly and the isotropic case is slightly higher (10\%) than the anisotropic one. 
The two straight lines are plotted using \eqref{mu01}.
$\mu_2$ and $\mu_3$ increase super-linearly and the anisotropic case is higher by 20\% and 35\% than the isotropic one at $\rho=4$ respectively.
For $\mu_2$, the two curves are obtained by numerically integrating \eqref{mu2}, and for $\mu_3$ they are fitted to $a-b\rho^{1/3}+c\rho$ for constants $a,b,c>0$

The middle and right panels of Fig. \ref{fig:iso} shows in different colors the gain (isotropic or anisotropic) which maximizes $\mu_k$ in the parameter space of $(\rho,\eta)$ for $k=1,2$.
Looking at the middle panel of Fig. \ref{fig:iso}, it is clear that for $\eta>2$ networks with isotropic antennas will typically have a higher $1$-hop mean degree than with anisotropic radiation patterns.
Significantly, this characterization is independent of the density $\rho$.
For $k=2$ however, anisotropic patterns are superior and occupy more and more space at higher node densities.
We therefore reach the following non-trivial conclusion: \textit{multihop accessibility is intensified for directional antennas, especially at high node densities.}
We expect that the more directional $G$ is, the better its multi-hop accessibility will be in the dense regime.
The mechanism by which this occurs is simple: 
\textit{1) Fixed $\eta$:} as node density $\rho$ increases, the next hop node is more likely to be oriented in a better direction (than from the forwarding node), resulting in better performance (in terms of reachability) of networks with anisotropic antennas.
\textit{2) Fixed $\rho$:} high path loss exponent $\eta$ degrades the performance of both radiation patterns, but does so more for the anisotropic one. This is because concentrated beams are more likely to be obstructed in cluttered environments resulting in better performance of networks with isotropic antennas.

\begin{figure}[t]
\centering
\includegraphics[scale=0.21]{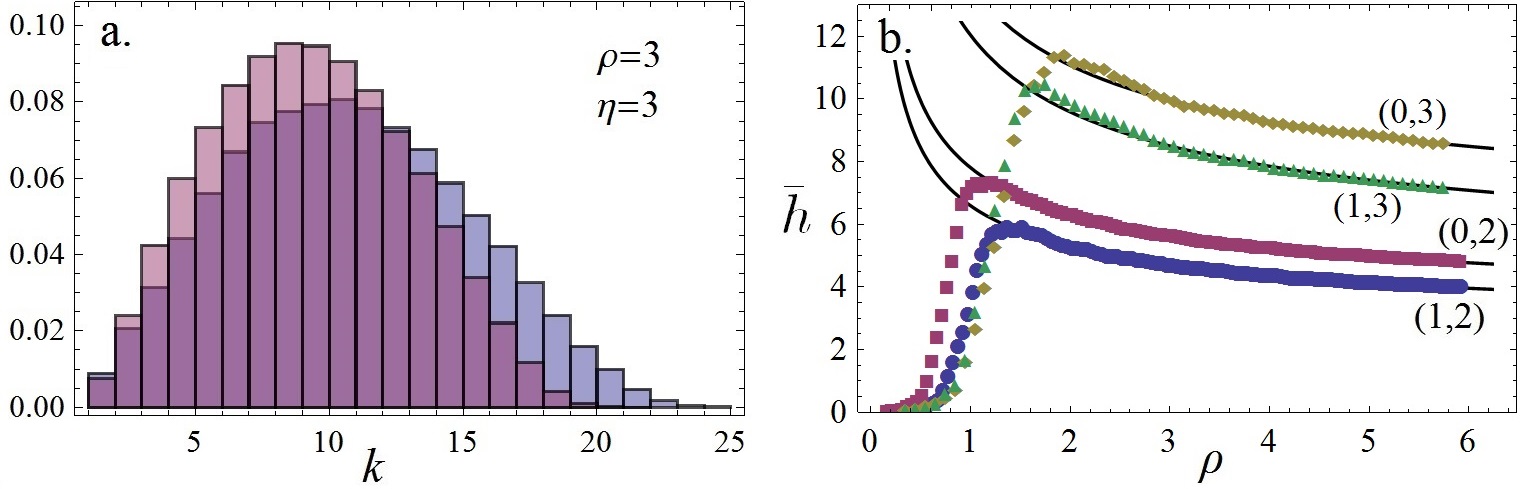}
\caption{\textit{Left:} Comparison of the hop distribution between isotropic (blue bars) and anisotropic (purple bars) networks using $\rho=\eta=3$. 
The $y$-axis therefore gives the probability that two randomly selected nodes are $k$-hops apart.
\textit{Right:} Plot showing the typical hop distance $\bar{h}$ as a function of the node density $\rho$ for different values of $(\epsilon, \eta)$ with fitted black curves for $\sim\rho^{-1/2}$.}
\label{fig:hopdist}
\end{figure}

Fig. \ref{fig:hopdist}a. shows the hop distribution obtained from a collection of networks generated using $\rho=\eta=3$ thus giving further insight into the comparison of reachability and network connectivity for isotropic and anisotropic antennas.
Recall that each bar in the histogram corresponds to the mean $k$-hop degree $\mu_k$.
As expected, $\mu_1$ is slightly larger for the isotropic case (blue) since $\eta>2$.
For $k>1$ the anisotropic distribution is skewed to the left indicative of the reachability benefits of networks with anisotropic antennas.
Physically this means that the resulting network has shorter paths (in terms of hops) between pairs of nodes; a clear indicator of potential gains in end-to-end delay subject to efficiently designed MAC and routing protocols \cite{bachir2010mac}. 
Fig. \ref{fig:hopdist}b. shows plots of the typical hop distance $\bar{h}(\rho)$ for different values of $(\epsilon, \eta)$.
$\bar{h}$ rapidly increases to a maximum at which point the network is fully connected, and then slowly decays like $\sim \rho^{-1/2}$, thus confirming Chandler's conclusion \cite{chandler1989calculation}.
It is clear however that the typical hop distance $\bar{h}$ is about 15\% less in mesh networks with randomly oriented directional antennas.

%%%%%%%%%%%%%%%%%%%%%
\section{Conclusions and Discussion \label{sec:conc}}

In this letter we have studied the multihop connectivity of networks with directional antennas. 
Specifically we have focused on random beamforming strategies where nodes choose the direction of radiation at random; a practical solution when no \textit{a priori} node location information is available or in random uncontrolled (e.g. air-dropped) WSN deployments.
We have developed a general mathematical framework and investigated the interplay between node density $\rho$, antenna directivity $G$, and the pathloss exponent $\eta$.
To this end, we have shown that random beamforming strategies can substantially improve multihop coverage, reachability, and also reduce the typical hop distance between nodes, thus improving end-to-end delay and signalling overheads.
Significantly, we have obtained closed form expressions for the expected one- and two- hop mean degree thus enabling further design optimization of large wireless networks with randomly oriented directional antennas.
Furthermore, knowledge of such statistics can be used to make MAC and routing protocols \textit{smarter}. 
%We emphasize that our qualitative results are general and independent of the specific fading model used herein (Rayleigh), the cardiod radiation pattern \eqref{cc}, or the circular domain $\V$.
In the future, it would certainly be interesting to confirm our findings using packet-level network simulators, e.g. ns3, or even in real experimental test-bed deployments.
It would further be interesting to extend this theoretical work on randomly oriented directional antennas to interference limited wireless networks by means of stochastic geometry methods.

%%%%%%%%%%%%%%%%%%%%%%%%%%
%\section*{Acknowledgements}
%\addcontentsline{toc}{section}{Acknowledgment}
%The authors gratefully acknowledge helpful discussions with J. Coon, C Dettmann, N. Deguchi, and A. Ganesh. The authors would also like to thank the directors of the Toshiba Telecommunications Research Laboratory for their support.

%%%%%%%%%%%%%%%%%%%%%%%%%%
\appendices

\section{2-hop mean degree for the hard disk model \label{sec:ap1}}
\numberwithin{equation}{section}

Consider the hard disk connection morel given by 
$H_{ij}= 1 $ if $r_{ij}<r_0$ and $0$ otherwise and substitute it into \eqref{mu2} to get
\es{
\mu_2 = 2\pi \rho \int_{r_{0}}^{2 r_0} r_{ij} \pr{ 1-e^{-\rho A(r_{ij})}} \dd r_{ij}
\label{muap}}
where $A(r_{ij})= 2r_{0}^2 \arccos\frac{r_{ij}}{2r_{0}} - \frac{r_{ij}}{2}\sqrt{4 r_{0}^2 - r_{ij}^2}$ is the intersection area of two overlapping circles of radius $r_{0}$, a distance $r_{ij}$ apart.
As $A(r_{ij})$ appears in the exponential of \eqref{muap}, the dominant contribution to $\mu_2$ at high node densities will be due to a situation when there is little to no overlap between the communication regions of nodes $i$ and $j$ i.e. when $A(r_{ij})\ll 1$ and a common $1$-hop neighbour is unlikely.
We therefore expand $A(r_{ij})$ near $r_{ij}\approx 2 r_{0}$ and obtain 
$A(r_{ij}) \approx \frac{4\sqrt{r_{0}}}{3} (2r_{0} - r_{ij})^{3/2} $.
Substituting this into \eqref{muap} we get
\est{
\mu_2 &\approx 2\pi \rho \int_{r_{0}}^{2r_{0}} \!\! r_{ij} \pr{1- e^{-\rho \frac{4\sqrt{r_{0}}}{3} (2r_{0} - r_{ij})^{3/2} }} \dd r_{ij} \\
&= 3\rho\pi r_{0}^2 - 2\pi(2r_{0})^{2/3} \Gamma\pr{2/3} \pr{\rho/3}^{1/3} + \mathcal{O} (\rho^{-1/3})
,}
since $A(r_{ij}>2r_{0})=0$. 
In the 3D network case we get $\mu_2 \sim 28/3 \pi \rho r_{0}^3  - 8 \pi r_{0}^{3/2} \sqrt{2\rho}$.

%%%%%%%%%%%%%%%%%%%%%%
\bibliographystyle{ieeetr}
\bibliography{IEEEabrv,mybib}

\end{document}